\documentstyle[aps,preprint,prl,epsfig]{revtex}
\tightenlines 
\begin{document}
\draft \flushbottom


\title{How different Fermi surface maps emerge in
photoemission from Bi2212}

\author{M.C. Asensio$^{1,2,*}$, J. Avila$^{1,2}$, L. Roca$^{1,2}$,
  A. Tejeda$^{1,2,*}$, G.D. Gu$^3$, M. Lindroos$^{4,5}$,
  R.S. Markiewicz$^{5}$, and A. Bansil$^{5}$}

\address{$^1$ Instituto de Ciencias de Materiales de Madrid (CSIC),
28049 Cantoblanco, Madrid, Espa\~na. \\
$^2$ LURE, B\^{a}t. 209D, Universit\'{e} Paris-Sud, BP34,91898 Orsay,
France \\
$^3$ Physics Department, Bldg. 510B, Brookhaven National Laboratory,
Upton, NY 11975-5000. \\
$^4$ Institute of Physics, Tampere University of Technology,
  33101 Tampere, Finland. \\
$^5$  Department of Physics, Northeastern University, Boston, MA 02115
USA.}

\date{\today}
\maketitle
\begin{abstract}

We report angle-resolved photoemission spectra (ARPES) from the Fermi energy
($E_F$) over a large area of the ($k_x,k_y$) plane using 21.2 eV and 32 eV
photons in two distinct polarizations from an optimally doped single crystal
of Bi$_2$Sr$_2$CaCu$_2$O$_{8+\delta}$ (Bi2212), together with extensive
first-principles simulations of the ARPES intensities. The results display
a wide-ranging level of accord between theory and experiment and clarify
how myriad Fermi surface (FS)
maps emerge in ARPES under various experimental conditions. The
energy and polarization dependences of the ARPES matrix element help
disentangle primary contributions to the
spectrum due to the pristine lattice from those arising from modulations
of the underlying tetragonal symmetry and provide a route for
separating closely placed FS sheets in low dimensional materials.

\end{abstract}
]

Establishing the nature of the Fermi surface (FS) in the metallic
state--the shape, size, connectivity and the number of FS
sheets\cite{ZX2,ZX3,Des3,kordyuk} and the
character of states at and near the Fermi energy ($E_F$), is crucially
important for identifying the correct superconducting scenario in the
cuprates. In this connection, high-resolution angle-resolved photoemission
spectroscopy (ARPES) provides a unique window in
low dimensional materials. However, even under identical conditions
of doping, temperature and other physical parameters, the imprint of the
FS in ARPES often varies dramatically with the energy and polarization of
the incident photons. This article reports fundamental progress in
understanding this remarkable behavior by considering the example of Bi2212
which has played the role of a touchstone and a workhorse in ARPES studies
of the high-Tc's. By comparing ARPES measurements over a large area of the
FS with corresponding first-principles simulations, we show how myriad
tapestries of FS maps originate in photoemission, and how the ARPES
matrix element blends and highlights various parts of the electronic
spectrum
--pristine states, effects of 
modulations\cite{Aeb,Camp1,fret,sato,saini,cite5}, different
FS sheets and the like, depending upon the specifics of the photon energy
and polarization.

High quality single crystals of optimally doped Bi2212 ($T_c$=90 K)
were grown by the traveling solvent floating zone method.
The FS maps were produced by measuring the photointensity within a narrow
energy window at $E_F$ over the full 360$^\circ$ angular range in
the ($k_x,k_y$) plane at the SU8 high-resolution beamline of the Super-Aco
synchrotron ring at LURE. \cite{Avila,Mascaraque}
The data were collected on a regular mesh in polar ($\theta$)
and azimuthal angles ($\phi$) of steps $\Delta\theta =1^o$, $\Delta\phi
=1^o$.
Some measurements employed He-I light (unpolarized) from a He discharge
lamp.
The overall energy resolution varied from 12 meV to 60 meV depending on
measurement conditions.\cite{Avila2,Gallego}
Procedures to assure accurate sample alignment included a complete
set of photoelectron diffraction azimuthal and polar scans of the Bi core
level in the high-energy regime which were recorded insitu in order to
define
the ${\bar\Gamma}$ point and the main high symmetry directions.
\cite{Ascolani1,Ascolani2,Huttel}
Two different detection geometries -- referred to as `even' and `odd' --
have been employed to obtain the ARPES spectra.
In the even case, the detector is constrained to move in the horizontal
plane,
which is the {\it same} as the plane of the incident light, whereas for the
odd case, the detector is chosen to move in the vertical plane.
The polarization is defined as measured with respect
to radial lines through $\Gamma$: even (odd) polarization implies the
polarization vector lies along (transverse to) this radial direction.
It can then be shown that
when emission from a mirror plane in the crystal lattice
is involved (e.g. along the $\Gamma - M$ line), the even (or odd)
experiment
only probes initial states of even (or odd) symmetry.
In the case of emission from the Cu-O
planes in Bi2212, the relevant initial states are predominantly Cu
$3d_{x^2-y^2}
$ hybridized with O $2p_{x,y}$ and
states along the ${\bar\Gamma}\rightarrow\bar M$ line (along the Cu-O-Cu
bonds) are expected to be intense in the even experiment, while states along
$\bar\Gamma\rightarrow\bar X(\bar Y)$ will be strong in the odd
measurements (since deviations from tetragonal order are small).

All computations in this article are based on the one-step model of
photoemission extended to treat arbitrarily complex unit cell materials
\cite{matri,stanf,ncco,cucis,mlcape,prl}.  Effects of multiple scattering
and the ARPES matrix element \cite{matri} are thus included realistically
in the presence of a specific surface termination, taken here to be the
Bi-O layer. The finite lifetimes of the initial and final
states are incorporated by giving suitable imaginary parts to the
self-energies of the associated initial and final state propagators.
The crystal potential was obtained via a selfconsistent KKR procedure
\cite{kkr1,kkr2,kkr3} assuming a perfect tetragonal lattice which yielded
the well-known LDA-based band structure and FS of Bi2212.
The actual potential used in this work however has been modified slightly
such that the Bi-O pockets around the $\bar M$-point are lifted above $E_F$
to account for their absence in the experimental ARPES spectra. A number of
simulations have also been carried out where repulsive barriers were placed
between the two CuO$_2$ planes in the Ca layers in order to mimic
correlation
effects beyond the LDA which are expected to reduce the bilayer splitting.
Finally, in order to describe the effects of the orthorhombic modulation
and
the $\sim (1 \times 5)$ superstructure present in Bi2212, we have taken the
spectrum computed for the tetragonal case and superposed on this the same
spectrum with appropriate weights after suitable translations in the
reciprocal space. The weights estimated from relative intensities in the
experimental FS maps are summarized in Table 1 and
described in the text below.

Figure~\ref{fig:1} considers two sets of theoretical spectra at 21.2 eV in
the
even and odd detection geometries for the pristine tetragonal lattice, and
helps
set the stage for our discussion. The right hand Figs.~\ref{fig:1}(b,d)
refer to
simulations where the bilayer splitting has been artificially reduced to a
near zero value and both FS sheets are hole-like around the
$\bar X(\bar Y)$ symmetry point.  The case with a bilayer splitting of
$\sim$ 200 meV where the first FS sheet is hole-like and the 2nd sheet
is slightly electron like is depicted in the left hand
Figs.~\ref{fig:1}(a,c).
We see by comparing Figs.~\ref{fig:1}(a) and \ref{fig:1}(b) that there is a
striking increase in the intensity around the $\bar M$-point for even detection
due to the presence of the aforementioned 2nd underlying FS sheet. In fact,
we
have carried out numerous other simulations in which the size of the
bilayer
splitting and/or the position of $E_F$ around the $\bar M$-point was
varied;
these results show clearly that about the only way in which one can obtain
a large intensity around the $\bar M$-point in the even detection geometry
for 21.2 eV photons is when there is an additional band
present very close to $E_F$ around
the $\bar M$-point. This is a crucially important observation which we will
recall below in analyzing the experimental data. Fig.~\ref{fig:1} also
highlights the differences between the even and odd maps more generally.
The 2nd
FS sheet so prominent in the even map of Fig.~\ref{fig:1}(a) is essentially
`invisible' in the corresponding odd map of Fig.~\ref{fig:1}(c). Also, the
odd
map is complementary to the even map in the sense that regions of high
intensity
in one map are often replaced by those of low intensity in the other and
vice versa.

The determination of the FS topology of the underlying pristine phase in
Bi2212
has been complicated by the presence of secondary features tied to
superlattice
and shadow bands. The results of Table 1 offer insight into how a headway
can
be made in disentangling primary and secondary features by exploiting the
energy and polarization dependences of the ARPES matrix element.  We see
that in the 21.2 eV even geometry, no secondary feature displays a weight
greater than 20 $\%$ of the primary features, and therefore, this energy
and polarization is well-suited for delineating the primary FS. In sharp
contrast, the weights of secondary features are significantly greater at
32 eV in both the even and odd spectra as well as in the 21.2 eV odd
spectrum.
Interestingly, the shadow feature is most prominent in the 32 eV odd case,
and the second superlattice image is visible in 32 eV even map, albeit
weakly.

Although {\it absolute} intensities are not our focus, one aspect of these
is particularly germane to this discussion. The theoretically computed
maximum absolute ARPES intensities of the {\it primary} features are in
the ratios: 21.2 eV (even): 21.2 eV (odd): 32 eV (even): 32 eV
(odd) = 5:2:1.2:0.7, in reasonable accord with the corresponding
experimental
values of: 5:1.7:0.8:0.6, uncertainty in determining absolute experimental
intensity notwithstanding. Applying these ratios to the data of Table 1
[which
gives the relative weights of the secondary features], it is found that
the absolute intensities of various {\it secondary} features are
not all that different
at the two photon energies considered for either the odd or
the even maps. In other words, the primary emissions become highlighted
in the 21.2 eV even measurements not because the secondaries are weak,
but due to the primaries becoming much stronger.

Figures~\ref{fig:2} and \ref{fig:3} compare experimental and theoretical
spectra
at 21.2 eV and 32eV respectively.  While changes in the ARPES maps of the
FS
with photon energy and
polarization are striking, an excellent overall agreement with theory is
seen
in all cases. The unpolarized (He I) data (Fig.~\ref{fig:2}(a)) and theory
(Fig.~\ref{fig:2}(b)) both show a clear imprint of the large hole sheet
centered
around $\bar X$ or $\bar Y$, but relatively little trace of the 2nd FS
sheet
around $\bar M$.  However, under even polarized light
(Figs.~\ref{fig:2}(c,d))
this 2nd sheet comes alive giving intense emission around the $\bar
M$-points.
As emphasized already in connection with Fig.~\ref{fig:1} above, the
appearance
of this large intensity is an unmistakable signature of the presence
of a second band lying close to E$_F$ at $\bar M$.
As noted above, under these experimental conditions secondary
features are quite weak (Table 1) and can be ignored in the
analysis of Figs.~\ref{fig:2}(c,d).
We see thus how a systematic comparison between experimental and theoretical
ARPES maps shows that the 21 eV even detection geometry is well suited for
establishing unambiguously the presence of the second FS sheet with
negligible interference from umklapps and shadow bands.

Under odd polarization (Figs.~\ref{fig:2}(e,f)) secondary features develop
greater relative intensity, so the even and odd maps look quite different.
As expected, the emission around $\bar M$ from the
2nd sheet is suppressed in this polarization.
On the other hand, the trace of the FS sheets is reinforced along the
lattice
modulation direction via superposition with its Umklapp images, thereby
distorting the image of the hole sheet in the odd map
(Fig.~\ref{fig:2}(e)), so
that caution must be exercised in deducing physical parameters
(size and shape) from this spectrum.
The mix of primary and secondary features is quite different in the 32eV
data,
Fig.~\ref{fig:3}, even though some of the characteristics are
basically similar. For example, the 2nd FS sheet is emphasized in the
even polarization in Figs.~\ref{fig:3}(a,b), but suppressed in the odd case
of Figs.~\ref{fig:3}(c,d). Compared to 21.2 eV, there is generally a
greater
spectral weight at higher momenta; for example, the image of the 2nd sheet
is more or less equally intense in the 1st and 2nd BZ's in
Figs.~\ref{fig:3}(a,b), but in the 21.2 eV Figs.~\ref{fig:2}(c,d), the 2nd
BZ
imprint is quite weak in relation to that in the 1st BZ.  The combination
of the
two FS sheets with appropriately weighted secondary images yields two
nearly
parallel bands of intense emission which are oriented along the
($\pi, -\pi$)  direction in Fig.~\ref{fig:3} and appear quite striking.
Interestingly, in even polarization (Figs.~\ref{fig:2}(c,d), or
\ref{fig:3}(a,b)) the spectra show a remarkable line of essentially zero
intensity along the $\bar\Gamma\rightarrow\bar Y$ direction, as expected
for a mirror plane.  Along the perpendicular line $\bar\Gamma
\rightarrow\bar X$, there is clear residual intensity -- a signature of
superlattice effects which break the symmetry selection rules.

In conclusion, we demonstrate that dramatic differences in the observed FS
maps under various experimental conditions arise as a consequence of the
ARPES matrix element which gives the primary FS features (arising from the
pristine lattice) a weight which depends upon the energy and polarization
of light in a {\it very different manner} from that of the secondary
features associated with the superlattice and orthorhombic modulations.
In this way the ARPES matrix element helps disentangle various aspects
of the electronic structure and fermiology of this complex system. In
particular, we confirm recent reports\cite{ZX3,Des3,kordyuk} of two 
FS sheets in
Bi2212,
extending the result to optimal doping.  There is one large hole-like standard
sheet and a second electron-like sheet which arises from a band which lies
very close the Fermi energy at the $\bar M$-point.
The wide ranging accord between theory and
experiment suggests that the band theory framework implicit in our
computations captures the essential underlying physics.
If so, the two FS pieces should be viewed as the result of conventional
bilayer splitting. However, it would be wise to keep the door open for
admitting other explanations, e.g. some form of a nano-scale phase
separation
or stripe order.

\acknowledgments

We acknowledge support by the Spanish agency DGES under
grant PB-97-1199, the U.S.D.O.E. Contracts W-31-109-ENG-38
and AC02-98CH10886, the large scale Installation Program
of the EU, and the allocation of supercomputer time
at the NERSC and the Northeastern University (NU-ASCC) Computation
Centers.

\begin{table}

\caption{
Relative fractional weights of various secondary features in Bi2212
obtained by analysing the experimental FS maps
where the primary spectrum for the
pristine lattice is normalized to unity. The superlattice modulation
gives the 1st and 2nd Umklapp images, while the orthorhombic modulation
yields the shadow feature and the related Umklapps. The translation
vectors involved are of form $\eta (\pi ,\pi )$ parallel to the
orthorhombic $b^*$ axis. Listed values of $\eta$ come in pairs.
1st and 2nd Umklapps roughly correspond
to $\eta$ values of $\pm 1/5$ and $\pm 2/5$, respectively.
}

\vspace*{0.5cm}

\begin{tabular}{|c|c||l|l|l|l|}
\tableline
$h\nu$ & detection& \multicolumn{2}{|c|}{orthorhombic}&\multicolumn{2}{|c|}
{superlattice}\\
& geometry& \multicolumn{2}{|c|}{modulation}&
\multicolumn{2}{|c|}{modulation}\\
\tableline
&& shadows&1$^{st}$ &$1^{st}$ &$2^{nd}$ \\
&&&umklapps&umklapps&umklapps\\
\tableline
21.2eV&even&0.10&0.01&0.20&0.01\\
&odd&0.30&0.15&0.40&0.02\\
\tableline
32 eV&even&0.15&0.05&0.35&0.12\\
&odd&0.40&0.12&0.40&0.01\\
    \tableline
\multicolumn{2}{|c|}{$\eta$ [Translation
}&$\pm$0.5&+(0.5$\pm$0.21)&$\pm$0.21&$\pm$0.42\\
\multicolumn{2}{|c|}{vector is $\eta(\pi ,\pi )$]}&&--(0.5$\pm$0.21)&&\\
    \end{tabular}
    \end{table}

\vfill\eject
\centerline{\bf Figure Captions}

\centerline{\bf Fig. 1}
Theoretical maps at 21.2 eV for emission from $E_F$ in
pristine (tetragonal) Bi2212 for even and odd polarizations
of light described in the text. The left hand simulations (a,c)
refer to the band theory-based crystal potential with a bilayer
splitting of $\sim$ 200 meV at the $\bar M$ point. In the right hand
side simulations (b,d), the bilayer splitting has been
artificially reduced to a nearly zero value.
Intensities are normalized to the same
value in all maps and plotted in hot colors (whites and yellows
are highs) on a linear scale. The square in solid white lines
marks the 2D Brillouin zone.

\centerline{\bf Fig. 2}
Theory and experiment are compared directly at 21.2 eV
for even and odd polarizations, as well as for the unpolarized
He I light. Effects of superlattice and orthorhombic
modulations are included in theory as discussed in the text. See caption to
Fig.~\protect\ref{fig:1} for other pertinent details.

\centerline{\bf Fig. 3}
Same as Fig.~\protect\ref{fig:2}, except that this figure refers to 32
eV synchrotron light.

\begin{figure}
\caption{Theoretical maps at 21.2 eV for emission from $E_F$ in
pristine (tetragonal) Bi2212 for even and odd polarizations
of light described in the text. The left hand simulations (a,c)
refer to the band theory-based crystal potential with a bilayer
splitting of $\sim$ 200 meV at the $\bar M$ point. In the right hand
side simulations (b,d), the bilayer splitting has been
artificially reduced to a nearly zero value.
Intensities are normalized to the same
value in all maps and plotted in hot colors (whites and yellows
are highs) on a linear scale. The square in solid white lines
marks the 2D Brillouin zone.}
\label{fig:1}
\end{figure}

\begin{figure}
\caption{Theory and experiment are compared directly at 21.2 eV
for even and odd polarizations, as well as for the unpolarized
He I light. Effects of superlattice and orthorhombic
modulations are included in theory as discussed in the text. See caption to
Fig.~\protect\ref{fig:1} for other pertinent details.}
\label{fig:2}
\end{figure}

\begin{figure}
\caption{Same as Fig.~\protect\ref{fig:2}, except that this figure refers
to 32
eV synchrotron light.}
\label{fig:3}
\end{figure}


\begin{references}
\bibitem {cite5}
We do not address the specific origin of the orthorhombic
features which could arise from various mechanisms, e.g. orthorhombic
lattice distortions, antiferromagnetic fluctuations, etc.
\end{references}
\end{document}